\documentclass[12pt, reqno]{article}
\usepackage{epsfig}
\usepackage{amssymb}
\usepackage{amsmath}
\usepackage{mathrsfs}
\usepackage{todonotes}
\usepackage{slashed}
\usepackage{hyperref}

\newcommand{\nin}{\in\!\!\! /}

\newcommand{\be}{\begin{eqnarray}}
\newcommand{\ee}{\end{eqnarray}}

\newcommand{\ab}[1]{\langle #1\rangle}

\begin{document}

\baselineskip=18pt

\setcounter{footnote}{0}
\setcounter{figure}{0}
\setcounter{table}{0}

\begin{titlepage}

\begin{center}

{\Large \bf A ``Twistor String" Inspired Formula For Tree-Level Scattering Amplitudes in ${\cal N}=8$ SUGRA}

\vspace{1.0cm}

{\bf Freddy Cachazo$^{a}$ and Yvonne Geyer$^{a,b}$}

\vspace{.3cm}

{\it $^{a}$ Perimeter Institute for Theoretical Physics, Waterloo, Ontario N2J W29, CA}

{\it $^{b}$ Physics Department, University of Waterloo, Waterloo, Ontario N2L 3G1, CA}

\vspace{0.5cm}

\end{center}

\begin{abstract}

We propose a new formulation of the complete tree-level S-matrix of ${\cal N}=8$ supergravity. The new formula for $n$ particles in the $k$ R-charge sector is an integral over the Grassmannian $G(2,n)$ and uses the Veronese map into $G(k,n)$. The image of a point in $G(2,n)$ is required to be in the ``complement" of a $2|8$-plane thus making the $SU(8)$ R-symmetry manifest. The integrand is the ratio of two determinants. The numerator is an analog of Hodges' recent determinant formula for MHV amplitudes. The denominator is a $2(n+k-2)\times 2(n+k-2)$ minor of a $2(n+k)\times 2(n+k)$ matrix of rank $2(n+k-2)$. Just as Hodges' formula does for MHV amplitudes, our integrand makes the complete invariance under $S_n$ manifest for all sectors.

The validity of the new formula follows from two surprising facts. One is the equivalence of Hodges' MHV formula and the Kawai-Lewellen-Tye (KLT) formula when kinematic invariants are allowed to be off-shell in a novel way. We give a proof of this for any number of particles. The second fact is an orthogonality property of the solutions to the polynomial equations defining the Veronese embedding. Explicit proof of the orthogonality is given for all amplitudes in all R-charge sectors with eight or less particles thus providing non-trivial evidence for our proposal.

\end{abstract}

\bigskip
\bigskip

\end{titlepage}

\section{Introduction and The Proposal}

Since the work of Witten in 2003 \cite{Witten:2003nn}, a very fruitful stream of new ideas has changed the way we think about and compute scattering amplitudes in gauge theory. Progress in the maximally supersymmetric Yang-Mills theory has been so dramatic that it would have been impossible to foresee in the early years following Witten's original work \cite{nice}.

Several early indications led some authors to believe that the gravity theory with maximal supersymmetry in four dimensions could actually be even more constrained and hence simpler than its gauge theory counterpart \cite{ArkaniHamed:2008gz}. This suspicion was supported by several arguments most involving the better behavior under Britto-Cachazo-Feng-Witten (BCFW) deformations \cite{Britto:2005fq, ArkaniHamed:2008yf}.

An overarching question regarding the simplicity of the supergravity theory is that of its tree level S-matrix. At this point all known formulations imply a large complexity due to the factorial growth in the number of terms. Such a factorial growth is also present in Yang-Mills amplitudes but it is tamed by the splitting into partial amplitudes \cite{Dixon:1996wi}. Of course, the total amplitude is a sum over at least $(n-3)!$ partial amplitudes weighted by color factors \cite{Bern:2008qj, Feng:2011gc}.

In ${\cal N}=4$ super Yang-Mills, very powerful techniques involving the geometry of Grassmannians \cite{ArkaniHamed:2009dn, Mason:2009qx} have led to remarkable progress not only at tree level but to all orders in perturbation theory \cite{ArkaniHamed:2010kv}. The Grassmannian formulation of Arkani-Hamed, Cheung, Cachazo and Kaplan (ACCK) makes all properties of individual partial amplitudes in Yang-Mills manifest, including invariance under the Yangian of the superconformal algebra \cite{ArkaniHamed:2009dn, ArkaniHamed:2009vw}. However, properties among partial amplitudes known as Kleiss-Kuijf (KK) \cite{Kleiss:1988ne} and Bern-Carrasco-Johansson (BCJ) \cite{Bern:2008qj} relations are obscure in the ACCK formulation. Fortunately, there is another formulation, derived from Witten's twistor string construction \cite{Witten:2003nn} by Roiban, Spradlin and Volovich (RSVW) \cite{Roiban:2004yf} which turns all these relations into simple-to-prove statements \cite{Roiban:2004yf, newpaper}.

In this paper we propose a formula for the full tree-level S-matrix of ${\cal N}=8$ supergravity which is the analog of the RSVW formula for ${\cal N}=4$ super Yang-Mills. The number of terms in the ${\cal N}=8$ supergravity formula grows only as a power of the number of particles. Comparing with the $(n-3)!$ growth in the RSVW Yang-Mills formulation, our formula strengthens the simplicity idea proposed in \cite{ArkaniHamed:2008gz}.

In order to establish conventions in the more familiar setting of ${\cal N}=4$ SYM, let us write the RSVW formula for a partial amplitude in the $k$ R-charge sector as an integral over the Grassmannian $G(2,n)$ using a degree $k-1$ Veronese map\footnote{For details on the connection to the original RSV formula see \cite{ArkaniHamed:2009dg}},
\be
\label{rsv}
{\cal A}^{(k)}(1,2,\ldots , n) =\frac{1}{{\rm vol}(GL(2))}\int \frac{d^{2n}\sigma}{(1~2)(2~3)\ldots (n~1)}\prod_{\alpha=1}^k\delta^{2}\left(\sum_{a=1}^n C^V_{\alpha ,a}(\sigma )\tilde\lambda_a \right)\times \nonumber \\ \prod_{\alpha=1}^k \delta^{0|4}\left(\sum_{a=1}^n C^V_{\alpha ,a}(\sigma )\tilde\eta_a \right)\int d^{2k}\rho \prod_{a=1}^n\delta^2\left( \sum_{\alpha=1}^k \rho_\alpha C^V_{\alpha , a}(\sigma )-\lambda_a\right).
\ee
Here and in the rest of the paper $(a~b)$ denotes a $2\times 2$ minor of
\be
\label{sigma}
\left(
  \begin{array}{cccc}
    \sigma^{(1)}_1 & \sigma^{(2)}_1 & \cdots & \sigma^{(n)}_1 \\
    \sigma^{(1)}_2 & \sigma^{(2)}_2 & \cdots & \sigma^{(n)}_2 \\
  \end{array}
\right).
\ee
The Veronese embedding of $G(2,n)$ into $G(k,n)$ is defined by
\be
C^V_{\alpha ,a}(\sigma ) = (\sigma_1^{(a)})^{k-\alpha}(\sigma_2^{(a)})^{\alpha -1}.
\ee

Now we have all the elements to define the proposal for the $n$ particle ${\cal N}=8$ supergravity amplitude in the $k$ R-charge sector
\be
\label{newf}
{\cal M}_{n,k} = \frac{1}{{\rm vol}(GL(2))}\int d^{2n}\sigma\int d^{2k}\rho \; \frac{H_n}{J_n} \; \prod_{\alpha=1}^k\delta^{2}\left(\sum_{a=1}^n C^V_{\alpha ,a}(\sigma )\tilde\lambda_a \right)\times \nonumber \\ \delta^{0|8}\left(\sum_{a=1}^n C^V_{\alpha ,a}(\sigma )\tilde\eta_a \right) \prod_{a=1}^n\delta^2\left( \sum_{\alpha=1}^k \rho_\alpha C^V_{\alpha , a}(\sigma )-\lambda_a\right)
\ee
where $H_n$ is computed from the $n\times n$ matrix $\Phi$
\be
\label{andrew}
\Phi_{ab} = \left\{
         \begin{array}{cc}
           \frac{s_{ab}}{(a~b)^2} & {\rm for} \;\; a\neq b \\
           -\sum_{c=1, c\neq a}^n \frac{s_{ac}}{(a~c)^2}\frac{(c~\ell)(c~r)}{(a~\ell)(a~r)}  & {\rm for} \;\; a=b \\
         \end{array}
       \right.
\ee
by deleting any three rows, say $\{a,b,c \}$, and any three columns, say $\{d,e,f\}$ and computing the determinant of the resulting matrix, denoted by $\Phi^{(abc)}_{(def)}$. The precise formula is
\be
\label{hfun}
H_n = \frac{1}{(a~b)(b~c)(c~a)}\times \frac{1}{(d~e)(e~f)(f~d)} |\Phi^{(abc)}_{(def)}|.
\ee
Note that in (\ref{andrew}), $(\bullet~ r)$ and $(\bullet ~ \ell)$ represent determinants with arbitrary reference $2$-vectors $r$ and $\ell$. The independence of $H_n$ on the choice of reference vectors and columns and rows to delete is simple and is shown in section 2. This matrix is very reminiscent of the recent MHV formula introduced by Hodges \cite{Hodges:2012ym} and we will see in section 4 how the two are directly connected.

Finally, the definition of $J_n$ is very similar to that of $H_n$. Here we have to define a $2(n+k)\times 2(n+k)$ matrix. In order to do so let us define two $2(n+k)$ dimensional vectors
\be
{\cal V} = \{ \rho_{1,1},\rho_{1,2},\ldots ,\rho_{k,1},\rho_{k,2}, \sigma_1^{(1)},\sigma_2^{(1)},\ldots ,\sigma_1^{(n)},\sigma_2^{(n)} \}
\ee
and
\be
{\cal E} = \{ E_{1,1},E_{1,2},\ldots ,E_{k,1},E_{k,2},F_{1,1},F_{1,2},\ldots ,F_{n,1},F_{n,2} \}
\ee
with
\be
E_{\alpha,\dot{\underline{\alpha}}} = \sum_{a=1}^n C^V_{\alpha ,a}(\sigma )\tilde\lambda_{a,\dot{\underline{\alpha}}},\quad F_{a,\underline{\alpha}} =\sum_{\alpha=1}^k \rho_{\alpha,\underline{\alpha}} C^V_{\alpha , a}(\sigma ).
\ee
The master $2(n+k)\times 2(n+k)$ matrix is then
\be
K_{I,J} = \frac{\partial {\cal E}_I}{\partial {\cal V}_J}.
\ee
On the support of the delta functions in (\ref{newf}) $K$ has rank $2(n+k-2)$. Eliminating any four rows corresponding to elements in $\cal V$ of the form $\{ \sigma_1^{(a)},\sigma_2^{(a)},\sigma_1^{(b)},\sigma_2^{(b)}\}$ and four columns corresponding to elements in ${\cal E}$ of the form $\{ F_{c,1},F_{c,2},F_{d,1},F_{d,2}\}$, we get a minor with non-vanishing determinant, denoted $K^{ab}_{cd}$. Just as in the previous case, it is possible to show that the following combination
\be
J_n = \frac{1}{(a~b)^2[c~d]^2} |K^{ab}_{cd}|
\ee
is independent of the choice of $\{ a,b,c,d\}$ and hence $S_n$ symmetric.

As we will see, this formula can be derived from a combination of the Kawai-Lewellen-Tye (KLT) relations \cite{Kawai:1985xq}, the RSVW construction and the determinant formula for MHV amplitudes introduced by Hodges in \cite{Hodges:2012ym}. The derivation, however, assumes the validity of two somewhat unexpected facts.

The first one is the equivalence of Hodges' MHV formula and the KLT MHV formula when the Lorentz invariant inner products of negative chirality spinors $[a~b]$ are replaced by entries $x_{ab}$ of an almost generic antisymmetric matrix. In particular, the $x_{ab}$ entries do not necessarily satisfy the Schouten identity. This is interesting as it can be interpreted as moving away from the $G(2,n)$ associated to the $\tilde\lambda$ variables in the Plucker embedding space. We give a complete proof of this fact in section 2.

The second fact is that the set of residues of the RSVW formula are orthogonal with respect to a bilinear form obtained from the KLT formula. This fact is the key to showing how the $SU(4)\times SU(4)$ R-symmetry which is manifest in KLT is enhanced to the full $SU(8)$ R-symmetry of supergravity.

In section 6, we show numerically that the orthogonality property of RSVW residues is satisfied for all amplitudes with eight or less particle in all R-charge sectors. This is a very nontrivial fact which reduces the number of terms for $n=6$, $n=7$ and $n=8$ and $k=3$ from the naive $4^2 =16$, $11^2=121$ and $26^2=676$ terms to only the diagonal, i.e., $4$, $11$ and $26$ respectively. For $n=8$ and $k=4$ one goes down from $66^2 = 4356$ to just $66$. The orthogonality property actually goes beyond individual R-charge sectors, we show that all RSVW residues for a given number of particles are orthogonal in our examples. Given that the number of RSVW residues for a given $n$ and $k$ is believed to be given by the Eulearian numbers $\left\langle \!\! \begin{array}{c}
n-3 \\ k-2 \end{array} \!\! \right\rangle$, their sum over $k$ is $(n-3)!$. This is also the size of the KLT bilinear form in the BCJ basis of partial amplitudes. This means that the RSVW solutions form a complete orthogonal basis with respect to the KLT bilinear.

The reader might want to skip sections 2 and 3 in a first reading and go directly to section 4 where the derivation of the formula is given. In section 5 we give the first steps towards a derivation of a $G(k,n)$ invariant formulation. Finally, in section 7 we start the exploration of several interesting lines of developments, like the possibility of defining an intrinsic off-shell MHV amplitude from correlators, a manifestly $S_n$ formulation obtained by introducing fermionic and bosonic redundancies and finally the twistor space form of the gravity formula.

\section{Hodges-KLT Generalized MHV Equivalence}

One of the two main ingredients in the derivation of the new formula for gravity amplitudes is the equivalence between Hodges' recent determinant formula for MHV gravity amplitudes \cite{Hodges:2012ym} and the KLT \cite{Kawai:1985xq} version when both amplitudes are continued off-shell in a novel way.

KLT is one of the first formulations that allowed the computation of all MHV gravity amplitudes. It relates gravity amplitudes to ``squares'' of partial amplitudes in Yang-Mills. This is one example for the factorial growth in the number of terms that gravity amplitudes have always been associated with, as one has to sum over permutations of $(n-3)$ elements.

Hodges shattered this expectation in his recent work \cite{Hodges:2012ym} by providing a formula for MHV gravity amplitudes as determinants of a $(n-3)\times (n-3)$ matrix, thus reducing the growth of terms from factorial to polynomial in $n$. This is therefore the formula of our choice for representing MHV amplitudes off-shell.

In the broader context, the proof of the equivalence between the off-shell continued version of Hodges formula and the KLT version will directly contribute to the derivation of the new formula. Starting from the KLT relations, the numerator of the integrand in the integral over $GL(2,n)$ contains the off-shell version of the KLT formula, the equivalence derived in this section allows for the rewriting of the numerator as $H_n$.

\subsection{Novel Off-Shell Deformation and Recursion Relations}

Let us start by describing what the new off-shell deformation corresponds to. Both Hodges and KLT MHV formulas are functions of the external data via the invariants $[a~b]$ and $\langle a~b\rangle$.

One can think of $[a~b]$ and $\langle a~b\rangle$ as entries of antisymmetric matrices $x_{ab}$ and $y_{ab}$ satisfying Schouten conditions
\be
\label{sch}
x_{ab}x_{cd} + x_{ad}x_{bc} + x_{ac}x_{db} = 0, \;\;\; y_{ab}y_{cd} + y_{ad}y_{bc} + y_{ac}y_{db} = 0
\ee
and momentum conservation
\be
\label{mom}
\sum_{b=1}^n x_{ab}y_{bc} = 0 \;\;\;\;\; \forall \;\{a,c\}.
\ee
We will leave completely untouched the $\langle a~b\rangle$ variables. The new off-shell deformation is to take $x_{ab}$ to be a general antisymmetric matrix which is only required to satisfy the momentum equation conditions (\ref{mom}). In other words, $x_{ab}$ will {\it not} generically satisfy the Schouten conditions (\ref{sch}). It is interesting to note that the Schouten conditions are precisely the Plucker equations for the embedding of a $G(2,n)$ Grassmannian into a projective space with coordinates $x_{ab}$. In recent years it has been useful to think about physical data as defining points in $G(2,n)$. Here we find that the off-shell deformation corresponds to moving away from the $G(2,n)$ subvariety in the projective space where it is embedded via the Plucker map.

Our strategy for proving the equivalence of both formulas in the off-shell setting is to show that both satisfy the same recursion relation. Given the fact that we have checked the equivalence for all $n<9$ using \texttt{Mathematica} this completes the proof. The recursion relation we use is the one presented by Hodges \cite{Hodges:2012ym}. Let use denote the ``generalized" amplitudes obtained from Hodges formula and from KLT as
\be
\Omega^{\rm Hodges}_n(\langle a~b\rangle, x_{ab}) \;\quad {\rm and} \;\quad \Omega^{\rm KLT}_n(\langle a~b\rangle, x_{ab})
\ee
respectively.

In our set up the recursion relation of Hodges has to be written as
\begin{equation}\label{eq:RecRel}
\begin{split}
 \Omega_n(\ab{a~b},x_{ab}&)\equiv\Omega_n(1,2,3,\dots,n\!-\!1,n)\\
 &=\sum_{p=3}^{n-1}\frac{x_{pn}}{\ab{p~n}}\frac{\ab{1~p}\ab{2~p}}{\ab{1~n}\ab{2~n}}\ \Omega_{n-1}(\hat{1},2,3,\dots,\hat{p},\dots,n\!-\!1)
\end{split}
\end{equation}
where the shifted variables are defined with respect to $p$ by
\begin{equation}
 x_{i\hat{1}}=x_{i1}+x_{in}\frac{\ab{n~p}}{\ab{1~p}}, \qquad \hat{\lambda}_1 =\lambda_1, \qquad x_{i\hat{p}}=x_{ip}+x_{in}\frac{\ab{n~1}}{\ab{p~1}}, \qquad \hat{\lambda}_p = \lambda_p.
\end{equation}

Summarizing, the goal of this section is to show that $\Omega^{\rm Hodges}_n(\langle a~b\rangle, x_{ab}) = \Omega^{\rm KLT}_n(\langle a~b\rangle, x_{ab})$.

\subsection{Hodges Generalized MHV Amplitude}

Hodges' original proof for the validity of his determinant formula reducing the number of terms from a factorial to a polynomial growth relies beautifully on the recursion relation (\ref{eq:RecRel}). In what follows, this proof will be generalized to allow the compatibility with the new off-shell deformation: As pointed out above, $x_{ab}$ will be taken to be a general antisymmetric matrix, just required to satisfy momentum conservation $\sum_{b}x_{ab}\ab{b~c}=0$, but not necessarily the Schouten identities (\ref{sch}).

\subsubsection{Definitions}

Let us start by recalling the definitions given in detail in \cite{Hodges:2012ym}: The $n\times n$ matrix $\Phi$, whose minors will provide the basic structure for the generalized MHV Hodges amplitude $\Omega^{\rm Hodges}_n$, is defined by
\begin{equation}
\begin{split}
 \phi^i_j&=\frac{x_{ij}}{\ab{i~j}}, \\
 \phi^i_i&=-\sum_{j\neq i}\frac{x_{ij}\ab{j~\ell}\ab{j~r}}{\ab{i~j}\ab{i~\ell}\ab{i~r}}
\end{split}
\end{equation}
where $\ell$ and $r$ are reference spinors. The matrix elements of $\Phi$ are independent of the choice of reference spinors, which can be seen straightforwardly, just relying on Schouten identities  of $\mathit{y_{ab}=\ab{a~b}}$ and momentum conservation.
For convenience, we will also introduce the factors $c_{ijk}=c^{ijk}=\frac{1}{\ab{i~j}\ab{j~k}\ab{k~i}}$. Having established these conventions, the generalized MHV Hodges amplitude is proposed to be
\begin{equation}\label{eq:Conj}
\Omega_n^{\rm Hodges}=(-1)^{n+1}\sigma(ijk,rst)c^{ijk}c_{rst}|\Phi|^{rst}_{ijk}
\end{equation}
where $\sigma(ijk,rst)=\text{sg}((ijk12\dots\slashed{i}\slashed{j}\slashed{k}\dots n)\rightarrow(rst12\dots\slashed{r}\slashed{s}
\slashed{t}\dots n))$ and $|\Phi|^{rst}_{ijk}$ denotes that minor of $\Phi$ where the rows $\{r,s,t\}$ and the columns $\{i,j,k\}$ have been removed.

\subsubsection{Recursion Relation}

Hodges showed that his determinant formula, obtained from our generalization by letting $x_{ab}$ be $[a~b]$, satisfies the recursion relation (\ref{eq:RecRel}).

The redundancy in the definition of the proposal (\ref{eq:Conj}) allows for a convenient choice of representation: The triple $\{12p\}$ for both the excluded rows and columns contains the minimal number of shifted variables. Combining the gravity amplitude for $\Omega_{n-1}$ with the recursion relation (\ref{eq:RecRel}) reduces the proof to
\begin{equation}\label{eq:Rec2}
 \Omega_n^{\rm Hodges}=\sum_{p=3}^{n-1}\frac{x_{pn}}{\ab{p~n}}\frac{\ab{1~p}\ab{2~p}}{\ab{1~n}\ab{2~n}} \Omega_{n-1}^{\rm Hodges} =(-1)^n\sum_{p=3}^n\phi^n_pc_{12n}c^{12p}|\widehat{\Phi}|^{12p}_{12p}.
\end{equation}
Here, $\widehat{\Phi}$ denotes an $(n\!-\!1)\times(n\!-\!1)$ matrix where the entries are defined with respect to the $n\!-\!1$ shifted momenta. The only difference occurs in the diagonal terms in the $p^{\text{th}}$ term of the sum:
\begin{align*}
 \widehat{\phi}^i_i&=-\sum_{j\neq i}\frac{x_{ij}\ab{j~\ell}\ab{j~r}}{\ab{i~j}\ab{i~\ell}\ab{i~r}}= \phi^{i}_i+\phi^n_i \frac{\ab{p~n}\ab{2~i}}{\ab{2~n}\ab{p~i}}.
\end{align*}
Following Hodges, for all further calculations, it will be useful to introduce a new matrix $F$, which two redundant rows and columns 1 and 2:
\begin{equation}
 f^i_j=\ab{1~i}\ab{2~j}\phi^i_j, \qquad \widehat{f}^i_j=\ab{1~i}\ab{2~j}\widehat{\phi}^i_j, \qquad \widehat{f}^i_i=f^i_i+f^n_i\frac{\ab{p~n}\ab{2~i}}{\ab{2~n}\ab{p~i}}.
\end{equation}
Using this, the recursion relation (\ref{eq:Rec2}) can be rewritten as
\begin{equation}
  (-1)^n\prod_{k=3}^{n-1}\ab{1~k}\ab{2~k}\Omega_n=c_{12n}c^{12n}\sum_{p=3}^n f^n_p|\widehat{F}|^{12p}_{12p}.
\end{equation}
When expanding the term $f^n_p|\widehat{F}|^p_p$ in the shift-correction parameters $f^n_{p_i}$, the 0$^{\text{th}}$ order correction is just $f^n_p|F|^p_p$, and the first order contributes with
\begin{align*}
 \sum_{3\leqslant p<q\leqslant n-1}&f^n_pf^n_q|F|^{pq}_{pq}\frac{\ab{p~n}\ab{2~q}}{\ab{2~n}\ab{p~q}}+f^n_qf^n_p|F|^{qp}_{qp}\frac{\ab{q~n}\ab{2~p}}{\ab{2~n}\ab{q~p}}
  =\sum_{3\leqslant p<q\leqslant n-1}f^n_pf^n_q|F|^{pq}_{pq}.
\end{align*}
The equality $1=\frac{\ab{p~n}\ab{2~q}}{\ab{2~n}\ab{p~q}}+(p\leftrightarrow q)$ used here generalizes in higher orders to
\begin{equation}\label{eq:idHodges}
 1=\sum_{j}\prod_{i\neq j}^m \frac{\ab{p_j~n}\ab{2p_i}}{\ab{2~n}\ab{p_j~p_i}},
\end{equation}
which is proven by Hodges and it clearly works in our set up as it only depends on $\ab{a~b}$.

The identity (\ref{eq:idHodges}) implies that the $i^{\text{th}}$ term contributes $\sum_{3\leqslant p_1<p_2<\dots<p_i\leqslant n-1}f^{n}_{p_1}\dots f^n_{p_i}|F|^{p_1p_2\dots p_i}_{p_1p_2\dots p_i}$ to the expansion, and thus the recursion relation simplifies further to
\begin{equation}
 (-1)^n\prod_{k=3}^{n-1}(1k)(2k)\Omega_n^{\rm Hodges}=c_{12n}c^{12n}\sum_{i=1}^3\sum_{3\leqslant p_1<p_2<\dots<p_i\leqslant n-1}f^{n}_{p_1}\dots f^n_{p_i}|F|^{p_1p_2\dots p_i}_{p_1p_2\dots p_i}.
\end{equation}
To prove this, Hodges uses a particularly beautiful trick: Consider an $(n\!-\!3)\times(n\!-\!3)$ matrix $H$ with entries
\begin{equation*}
 h^i_j=f^i_j+\delta^i_jf^n_j\quad \text{for} \;\; 3\leqslant i,j\leqslant n-1.
\end{equation*}
Each row of $H$ sums to zero due to $\sum_{i=1}^n f^i_n=0$, and hence the determinant of $H$ vanishes.
On the other hand, $|H|$ can be expanded in the shift variables $f^n_i$ similar to $|F|$. Again, the $i^{\text{th}}$ order contributes with $\sum_{3\leqslant p_1<p_2<\dots<p_i\leqslant n-1}f^{n}_{p_1}\dots f^n_{p_i}|F|^{p_1p_2\dots p_i}_{p_1p_2\dots p_i}$, therefore the determinant becomes
\begin{equation*}
 0=|H|=|F|+\sum_{3\leqslant p_1<p_2<\dots<p_i\leqslant n-1}f^{n}_{p_1}\dots f^n_{p_i}|F|^{p_1p_2\dots p_i}_{p_1p_2\dots p_i}
\end{equation*}
and hence the recursion relation reduces further to
\begin{equation}
 (-1)^n\prod_{k=3}^{n-1}\ab{1~k}\ab{2~k}\Omega_n^{\rm Hodges}=c_{12n}c^{12n}|F|.
\end{equation}
This is trivially proven by inserting the proposal (\ref{eq:Conj}), $\Omega_n^{\rm Hodges}=(-1)^{n+1}c_{12n}c^{12n}|\Phi|^{12n}_{12n}$, thereby concluding the proof that $\Omega^{\text{Hodges}}(\ab{a~b},x_{ab})$ satisfies the recursion relation. \hfill$\Box$

\subsection{KLT Generalized MHV Amplitude}

The last step in the proof of the equivalence $\Omega^{\rm Hodges}_n(\langle a~b\rangle, x_{ab}) = \Omega^{\rm KLT}_n(\langle a~b\rangle, x_{ab})$ is to show that the generalized KLT amplitude satisfies the same recursion relation (\ref{eq:RecRel}).

Luckily, a proof that the KLT formula for the usual MHV amplitudes satisfies (\ref{eq:RecRel}) is reasonably simple and was done by Feng \cite{pricom}. Here we follow the same steps showing that nothing in the proof depends on the $x_{ab}$ satisfying the Schouten identity\footnote{We would like to thank Bo Feng for sharing with us his proof and allowing us to adapt it to our problem.}.

It turns out that the proof is slightly simpler if the role of particles $1$ and $n$ is exchanged. This can freely be done as KLT and Hodges' formula are permutation invariant. This means that the relabeled recursion relations are
\begin{equation}
\begin{split}
 \Omega^{\text{KLT}}(\ab{a~b},x_{ab}&) = \sum_{p=2}^{n-2}\frac{x_{pn-1}}{\ab{p~n-1}}\frac{\ab{1~p}\ab{n~p}}{\ab{1~n-1}\ab{n~n-1}}\ \Omega_{n-1}^{\rm KLT}(1,2,3,\dots,\hat{p},\dots,n\!-\!2,\hat n)
\end{split}
\end{equation}
where
\begin{equation}
 x_{i\hat{n}}=x_{in}+x_{in-1}\frac{\ab{n-1~p}}{\ab{n~p}},\qquad \hat{\lambda}_1=\lambda_1, \qquad x_{i\hat{p}}=x_{ip}+x_{in-1}\frac{\ab{n-1~n}}{\ab{p~n}},\qquad \hat{\lambda}_p=\lambda_p.
\end{equation}

The starting point is
\be
\Omega_n^{\rm KLT} = \sum_{p=2}^{n-2}\sum_{\alpha \in S_{n-4}}A^{\rm MHV}(1,\alpha ,n-1,n)\sum_{\tilde\beta \in S_{n-3}}{\cal S}[\tilde\beta |\alpha,p]_{p_1}A^{\rm MHV}(n,\beta ,1, n-1)
\ee
where ${\cal S}[\tilde\beta |\alpha,p]_{p_1}$ is only a function of $\ab{a~b}x_{ab}$. The precise form will not be relevant at this point (for a detailed definition see e.g. \cite{Feng:2011gc}).

One can split the set of permutations $\beta$ in groups where all elements except $p$ keep their relative ordering. After summing over all possible positions of $\beta$ one finds
\be
\Omega_n^{\rm KLT} = \sum_{p=2}^{n-2}\sum_{\alpha\in S_{n-4}}\!\!\! A^{\rm MHV}(1,n,\alpha ,p,n-1,n)\!\!\! \sum_{\beta \in S_{n-4}}\!\! {\cal S}[\beta |\alpha ]_{p_1}
\ab{n-1~p}x_{n-1p}A^{\rm MHV}(p,n,\beta,1,n-1).\nonumber
\ee
Using the Parke-Taylor formula for gauge theory MHV amplitudes $\Omega_n^{\rm KLT}$ becomes
\be
\sum_{p=2}^{n-2}\sum_{\alpha\in S_{n-4}}\!\! A^{\rm MHV}(1,n,\alpha ,p,n-1,n)\!\! \sum_{\beta \in S_{n-4}}\! {\cal S}[\beta |\alpha ]_{p_1}
\ab{n-1~p}x_{n-1p}A^{\rm MHV}(n,\beta,1,p,n-1)\times Q \nonumber
\ee
with
\be
Q = \frac{\ab{1~p}\ab{n-1~n}}{\ab{1~n-1}\ab{p~n}} .
\ee
Using the definition of the hatted variables momenta for $\hat{p}$ and $\hat{n}$ one finds
\be
\Omega_n^{\rm KLT} = & \sum_{p=2}^{n-2}\sum_{\alpha\in S_{n-4}}A^{\rm MHV}(1,\alpha ,\hat{p},\hat{n})\sum_{\beta \in S_{n-4}}{\cal S}[\beta |\alpha]_{p_1}A^{\rm MHV}(\hat{n} ,\beta,1,\hat{p})\times \tilde Q\nonumber
\ee
where
\be
\tilde Q = \frac{x_{pn-1}\ab{1~p}\ab{n~p}}{\ab{p~n-1}\ab{1~n-1}\ab{n~n-1}}.
\ee
Since the last factor does not depend on the permutations it can be factored out and the sum over the permutations precisely gives the amplitude with one less graviton, i.e.,
\be
\Omega_n^{\rm KLT} = \sum_{p=2}^{n-2}\left(\frac{x_{pn-1}\ab{1~p}\ab{n~p}}{\ab{p~n-1}\ab{1~n-1}\ab{n~n-1}}\right)\Omega_{n-1}^{\rm KLT}(1,2,\ldots ,\hat{p},\ldots , n-2,\hat{n}).
\ee

This concludes the proof of the equivalence between Hodges' and KLT MHV generalized formulas.

\section{The Orthogonality Conjecture}

The last ingredient to derive our formula for gravity amplitudes is a property that individual solutions to the RSVW equations are here conjectured to satisfy. The KLT formula naturally defines a bilinear form, $S_{\alpha,\beta}^{\rm KLT}$, acting on the vector space of partial amplitudes\footnote{In order to properly define the bilinear form one should use a BCJ basis of partial amplitudes on the left and on the right. In our discussion this will not be important but it might be the key for an analytic proof of the orthogonality conjecture.}. If partial amplitudes are replaced by the integrand of the RSVW formula \cite{Witten:2003nn,Roiban:2004yf}, i.e,
\be
\label{noti}
A_n(1,2,\ldots ,n) \to {\cal I}^{\rm RSVW}_{i} = \frac{1}{(1~2)_i(2~3)_i\ldots (n~1)_i}
\ee
evaluated at solution $I$, then the conjecture, in schematic form, is that
\be
\sum_{\alpha, \beta} {\cal I}^{\rm RSVW}_{i}(\alpha)\; S_{\alpha,\beta}^{\rm KLT}\;\; {\cal I}^{\rm RSVW}_j(\beta ) = 0 \;\; {\rm whenever} \;\; i\neq j.
\ee
It is important to mention that for a given number of particles, the conjecture refers to all solutions, including all different R-charge sectors. Clearly the orthogonality property is known to be satisfied for full partial amplitudes in different $k$ charge sectors \cite{Feng:2010br,Damgaard:2012fb}. This means that the conjecture is only about the behavior of individual RSVW solutions including distinct solutions in the same $R$-charge sector.

In section 6, we provide evidence for this conjecture by showing numerically that it is valid for all solutions with eight or less particles.

\subsection{The RSVW Integrand}

In the $G(2,n)$ formulation of the RSVW construction for Yang-Mills amplitudes one constructs a $2\times n$ matrix
\be
\left(
  \begin{array}{cccc}
    \sigma^{(1)}_1 & \sigma^{(2)}_1 & \cdots & \sigma^{(n)}_1 \\
    \sigma^{(1)}_2 & \sigma^{(2)}_2 & \cdots & \sigma^{(n)}_2 \\
  \end{array}
\right)
\ee
representing a point in $G(2,n)$. There is a $GL(2)$ action on the left which can be used to fix the value of four entries. Therefore, the matrix has $2n-4$ degrees of freedom.

We are interested in points in $G(2,n)$ that satisfy the RSVW equations reviewed in the introduction. They are
\be
\label{rsveq}
\sum_{a=1}^n C^V_{\alpha ,a}(\sigma )\tilde\lambda_{a,\dot{\underline{\alpha}}} = 0,\quad \sum_{\alpha=1}^k \rho_{\alpha,\underline{\alpha}} C^V_{\alpha , a}(\sigma ) = \lambda_{a,\underline{\alpha}}.
\ee
where
\be
C^V_{\alpha ,a}(\sigma ) = (\sigma_1^{(a)})^{k-\alpha}(\sigma_2^{(a)})^{\alpha -1}
\ee
and $\rho_{\alpha,\underline{\alpha}}$ is a $2\times k$ matrix of unknowns. The number of equations is $2(n+k)$ while that of variables is $2(n+k-2)$. This is not a problem if we assume that the external data satisfy momentum conservation as this reduces the rank of the system of equations by 4.

The equations (\ref{rsveq}) are polynomial equations and generically have many different solutions for a fixed $k$. Let us denote the number of solutions by ${\cal N}_{k,n}$.

Let us also refine even further the notation introduced in (\ref{noti})
\be
{\cal I}^{(k)}_i(1,2\ldots ,n) = \left.\frac{1}{(1~2)(2~3)\ldots (n~1)}\right|_{i^{th} \; {\rm solution}}
\ee
for the integrand of the RSVW formula evaluated on the $i^{\rm th}$ solution in the $k$ R-charge sector.

\subsection{Precise Formulation of the Conjecture}

The precise form of the conjecture states that for any $i\neq j$ and $k_L=k_R$ or $k_L\neq k_R$ the following holds
\begin{equation}\label{eq:KLTfull}
\begin{split}
 &\sum_{\sigma\in S_{n-3}}{\cal I}^{(k_L)}_i(1,\sigma(I\cup J),n-1,n)\times\\
 &\hspace{20pt}\times\sum_{\alpha\in S_\mu}\sum_{\beta\in S_\nu}f\big(\alpha\circ\sigma(I)\big)\overline{f}\big(\beta\circ\sigma(J)\big){\cal I}_j^{(k_R)}\big(\alpha\circ\sigma(I),1,n-1,\beta\circ\sigma(J),n\big) = 0
\end{split}
\end{equation}
where $I=\{i_1,\dots,i_\mu\}\in \mathcal{P}(2,\dots,\frac{n}{2})$ and $J=\{j_1,\dots,j_\nu\}\in \mathcal{P}(\frac{n}{2}+1\dots n-2)$ and where \footnote{Note that the definitions for the functions $f$ ad $\overline{f}$ are to be understood as evaluated for the ordering $\sigma(I)$ and $\sigma(J)$ respectively. This becomes relevant in the definition of $g$.}
\begin{align}\label{eq:ff}
 f(I)=f(i_1,\dots,i_\mu)&=s_{1,i_\mu}\prod_{m=1}^{\mu-1}(s_{1,i_m}+\sum_{k=m+1}^\mu g(i_m,i_k))\\
 \overline{f}(J)=\overline{f}(j_1,\dots,j_\nu)&=s_{j_1,n-1}\prod_{m=2}^{\nu}(s_{j_m,n-1}+\sum_{k=1}^{m-1} g(j_k,j_m))
\end{align}
with \hspace{80pt}$g(i,j)=\begin{cases} &s_{i,j}\quad i>j\\&0\quad \text{otherwise}.\end{cases}$\\

As will be discussed below, this conjecture establishes the symmetry enhancement from the manifest $SU(4)\times SU(4)$ R-symmetry of the KLT relations to the $SU(8)$ R-symmetry of $\mathcal{N}=8$ SUGRA as a trivial corollary, since the fermionic $\delta$-functions in the RSV formula combine naturally when evaluated on the same solution. This conjecture also implies a dramatic reduction in the number of terms as naively one would have to sum over ${\cal N}_{k,n}^2$ terms and now it is only ${\cal N}_{k,n}$.

\section{Derivation of the New Formula}

Having introduced all the ingredients we now proceed to derive the new formula. The starting point is the fully supersymmetric version of KLT that relates partial amplitudes in ${\cal N}=4$ SYM with amplitudes in ${\cal N}=8$ supergravity \cite{Feng:2010br,Damgaard:2012fb}.

There are several equivalent formulations but the one we will use here is \cite{Bern:1998sv}
\begin{equation}
\begin{split}
 M_n= &\sum_{\sigma\in S_{n-3}}A_n^{(L)}(1,\sigma(I\cup J),n-1,n)\times\\
 &\hspace{20pt}\times\sum_{\alpha\in S_\mu}\sum_{\beta\in S_\nu}f\big(\alpha\circ\sigma(I)\big)\overline{f}\big(\beta\circ\sigma(J)\big)A_n^{(R)}\big(\alpha\circ\sigma(I),1,n-1,\beta\circ\sigma(J),n\big)
\end{split}
\end{equation}
where the meaning of all the symbols is identical to the formula (\ref{eq:KLTfull}) in the previous section.

Note that $A_n^{(L)}$ and $A_n^{(R)}$ in the expression above denote partial tree-level amplitudes in ${\cal N}=4$ SYM and $M_n$ is the ${\cal N}=8$ supergravity amplitudes all stripped off the momentum conserving delta functions.

\subsection{Inserting The RSVW Formulation}

Straightforwardly inserting the RSVW formula (\ref{rsv}) into the KLT expression clearly does not make sense. The RSVW formula computes partial amplitudes ${\cal A}_n$ that contain a momentum conserving delta function. If inserted in KLT directly one gets two copies of the momentum conserving delta function. One way out is to insert the full RSV formula for ${\cal A}_R$ and for the left we insert
\be
A_L(1,2,\ldots ,n) =\!\!\!\! \sum_{L\in {\rm RSV solutions}}\!\! \frac{1}{(1~2)_L(2~3)_L\ldots (n~1)_L}\times \frac{1}{J_n(\sigma_L ,\rho_L)}\prod_{\alpha=1}^k\delta^{0|4}\left(\sum_{a=1}^n C^V_{\alpha ,a}(\sigma_L )\tilde\eta^{(L)}_a \right).\nonumber
\ee
Here $J_n$ is the RSVW jacobian and we have indicated by a subscript $L$ on $\sigma$'s and $\rho$'s that they are evaluated on the solutions of the RSVW equations coming from the ``left" amplitude. We have also introduced the notation $\tilde\eta^{(L)}$ to indicate the supersymmetric variables on the left amplitude.

Let us postpone the evaluation of the RSV jacobian and continue with the line of the argument. The only information we need at the moment is that the jacobian is independent of the chosen partial amplitude.

Applying these substitutions into the KLT formula we obtain a formula for the full gravity amplitude including a momentum conserving delta function. Indicating the structure of the permutations somewhat schematically, in order to avoid an impossible-to-read formula, we have
\be
{\cal M}_n = &\sum_{L\in {\rm RSV}}\left( \sum_{\gamma\in S_{n-3}}\frac{1}{(1~\gamma_2)_L(\gamma_2~\gamma_3)_L\ldots (\gamma_{n}~1)_L}\frac{1}{J_n(\sigma_L ,\rho_L)}\prod_{\alpha=1}^k\delta^{0|4}\left(\sum_{a=1}^n C^V_{\alpha ,a}(\sigma_L )\tilde\eta^{(L)}_a \right)\right)\times \nonumber \\ & \sum_{\alpha}\sum_{\beta}f\big(\alpha)\overline{f}\big(\beta)
 \frac{1}{{\rm vol}(GL(2))}\int \frac{d^{2n}\sigma_R}{(\alpha_1~\alpha_2)_R\ldots (1~n-1)_R\ldots (\beta_{n-1},n)_R}\times \nonumber \\ &\prod_{\alpha=1}^k\delta^{2}\left(\sum_{a=1}^n C^V_{\alpha ,a}(\sigma_R )\tilde\lambda_a \right)\delta^{0|4}\left(\sum_{a=1}^n C^V_{\alpha ,a}(\sigma_R )\tilde\eta^{(R)}_a \right)\times \nonumber \\ & \int d^{2k}\rho_R \prod_{a=1}^n\delta^2\left( \sum_{\alpha=1}^k \rho^{(R)}_\alpha C^V_{\alpha , a}(\sigma_R )-\lambda_a\right).
\ee
This formula looks very far from being or possibly leading to any kind of improvement. However, here is where the orthogonality property of RSVW residues comes to the rescue.

In the above formula for ${\cal M}_n$ we know that on the right part of the amplitude we are also supposed to solve the delta functions and sum over the solutions. The delta functions are exactly the same as for the ``left" problem. Therefore the values of the variables $\sigma_L, \rho_L$ and $\sigma_R$ and $\rho_R$ agree when evaluated on the same solutions. Recall that ${\cal N}_{k,n}$ denotes the number of solutions, then by multiplying the left and right terms we have a total of ${\cal N}_{k,n}^2$ combinations. Only in the diagonal terms, ${\cal N}_{k,n}$ of the total, are both $\sigma_L$ and $\sigma_R$ the same.

The reason these diagonal terms are very appealing is that the supersymmetric delta functions nicely combine
\be
\prod_{\alpha=1}^k\delta^{0|4}\left(\sum_{a=1}^n C^V_{\alpha ,a}(\sigma_L )\tilde\eta^{(L)}_a \right)\delta^{0|4}\left(\sum_{a=1}^n C^V_{\alpha ,a}(\sigma_R )\tilde\eta^{(R)}_a \right) = \prod_{\alpha=1}^k\delta^{0|8}\left(\sum_{a=1}^n C^V_{\alpha ,a}(\sigma_D )\tilde\eta_a \right)
\ee
where we have denoted by $\sigma_D$ the diagonal value, i.e., $\sigma_D = \sigma_R = \sigma_L$ and introduced
\be
\tilde\eta_a = \left(
                 \begin{array}{c}
                   \tilde\eta^{(R)}_a \\
                   \tilde\eta^{(L)}_a \\
                 \end{array}
               \right)
\ee
as an eight component Grassmann vector.

Clearly the orthogonality property discussed in section 3 precisely implies that all cross terms vanish individually and therefore we see directly the enhancement of the symmetry manifest in the formula from $SU(4)\times SU(4)$ to the full $SU(8)$ R-symmetry of ${\cal N}=8$ supergravity.

Not only the R-symmetry becomes manifest but we can also write a much simpler formula containing a single set of integration variables $\{\sigma ,\rho\}$, i.e., no more ``left" and ``right" subscripts are needed,
\be
& {\cal M}_{n,k} = \frac{1}{{\rm vol}(GL(2))}\int d^{2n}\sigma\int d^{2k}\rho \; \frac{1}{J_n(\sigma , \rho)}\times \nonumber\\
& \left(\sum_{\gamma\in S_{n-3}}\frac{1}{(1~\gamma_2)(\gamma_2~\gamma_3)\ldots (\gamma_{n}~1)}\sum_{\alpha}\sum_{\beta}f\big(\alpha)\overline{f}\big(\beta)
 \frac{1}{(\alpha_1~\alpha_2)\ldots (1~n-1)\ldots (\beta_{n-1},n)}\right)\nonumber\\
& \prod_{\alpha=1}^k\delta^{2}\left(\sum_{a=1}^n C^V_{\alpha ,a}(\sigma )\tilde\lambda_a \right)
\delta^{0|8}\left(\sum_{a=1}^n C^V_{\alpha ,a}(\sigma )\tilde\eta_a \right) \prod_{a=1}^n\delta^2\left( \sum_{\alpha=1}^k \rho_\alpha C^V_{\alpha , a}(\sigma )-\lambda_a\right).
\ee

\subsection{Simplifying the KLT Core}

At this point we are only one step away from our final form. In order to proceed we have to carefully study the part of the integrand given by
\be
\label{omg}
\sum_{\gamma\in S_{n-3}}\frac{1}{(1~\gamma_2)(\gamma_2~\gamma_3)\ldots (\gamma_{n}~1)}\sum_{\alpha}\sum_{\beta}f\big(\alpha)\overline{f}\big(\beta)
 \frac{1}{(\alpha_1~\alpha_2)\ldots (1~n-1)\ldots (\beta_{n-1},n)}.
\ee
Recall that the functions $f$ and $\bar f$ are given as products of polynomials in $s_{ab}$ with coefficients that are constants (see eq. \ref{eq:KLTfull}). The fact that $f$ and $\bar f$ do not depend on $\langle a~b\rangle$ and $[a~b]$ independently is the crucial property that will allow us to simplify the formula.

On the support of the delta functions we have that
\be
\lambda_a = \sum_{\alpha=1}^k \rho_\alpha C^V_{\alpha , a}(\sigma ).
\ee
This means that
\be
\label{mag}
\langle a~b\rangle = (a~b)P_{k-2}(\{\sigma_a,\rho_a\},\{\sigma_b,\rho_b\})
\ee
where $P_{k-2}$ is a homogeneous polynomial of degree $k-2$ in $\sigma_a$ and $\sigma_b$. Moreover, $P_{k-2}$ is symmetric under the exchange of the labels $(a)$ and $(b)$.

Using (\ref{mag}) to rewrite
\be
\label{ali}
s_{ab} = (a~b)x_{ab}\;\;\;\; {\rm with}\;\;\;\; x_{ab} = [a~b]P_{k-2}(\{\sigma_a,\rho_a\},\{\sigma_b,\rho_b\})
\ee
and plugging it into the definition of $f$ and $\bar f$ we find that (\ref{omg}) is identical to the function defined in section 2.3, i.e., the generalized or off-shell KLT MHV formula,
\be
\Omega^{\rm KLT}((a~b), x_{ab}).
\ee
Recall that this function becomes identical to $\Omega^{\rm Hodges}((a~b),x_{ab})$ if the variables $x_{ab}$ satisfy
\be
x_{ab} = -x_{ba}\;\;\;\; {\rm and} \;\;\;\; \sum_{b=1}^n x_{ab}(b~c) = 0 \;\;\;\; \forall \;\; \{a,c\}.
\ee
The antisymmetry property of $x_{ab}$ as defined in (\ref{ali}) is obvious. The second property is much more interesting. Using the definition we have
\be
\sum_{b=1}^n x_{ab}(b~c) = \sum_{b=1}^n [a~b]P_{k-2}(\{\sigma_a,\rho_a\},\{ \sigma_b,\rho_b \})(b~c).
\ee
Now we follow the same argument as the one used in \cite{newpaper} for the proof of the fundamental BCJ identities. On the support of the delta functions
\be
\label{pol}
\sum_{b=1}^n C^V_{\alpha ,b}(\sigma )\tilde\lambda_b = 0.
\ee
Recalling that $C^V_{\alpha ,b} = (\sigma^{(b)}_1)^{k-\alpha}(\sigma^{(b)}_2)^{\alpha -1}$ we can take linear combinations of (\ref{pol}) to produce any homogeneous polynomial $Q$ of degree $k-1$ in $\sigma_b$ and get
\be
\sum_{b=1}^n Q_{k-1}(\sigma_b)\tilde\lambda_b = 0.
\ee
Finally, note that
\be
P_{k-2}(\{\sigma_a,\rho_a\},\{ \sigma_b,\rho_b \})(b~c)
\ee
is a homogenous polynomial in $\sigma_b$ of degree $k-1$ and therefore
\be
\sum_{b=1}^n [a~b]P_{k-2}(\{\sigma_a,\rho_a\},\{ \sigma_b,\rho_b \})(b~c) =0
\ee
since it is multiplied with $\tilde\lambda_b$ and summed over $b$.

This concludes the derivation of our new formula if we use (\ref{mag}) to write $x_{ab} = s_{ab}/(a~b)$, then
\be
\Omega^{\rm Hodges}((a~b),x_{ab}=s_{ab}/(a~b)) = \frac{1}{(a~b)(b~c)(c~a)}\times \frac{1}{(d~e)(e~f)(f~d)} |\Phi^{(abc)}_{(def)}|
\ee
which is nothing but $H_n$ in (\ref{hfun}).

Summarizing, we have shown that starting with the KLT formula, appropriately inserting RSVW partial amplitudes, assuming the orthogonality conjecture and using the Hodges-KLT generalized equivalence one gets
\be
{\cal M}_{n,k} = \frac{1}{{\rm vol}(GL(2))}\int d^{2n}\sigma\int d^{2k}\rho \; \frac{H_n}{J_n} \; \prod_{\alpha=1}^k\delta^{2}\left(\sum_{a=1}^n C^V_{\alpha ,a}(\sigma )\tilde\lambda_a \right)\times \nonumber \\ \delta^{0|8}\left(\sum_{a=1}^n C^V_{\alpha ,a}(\sigma )\tilde\eta_a \right) \prod_{a=1}^n\delta^2\left( \sum_{\alpha=1}^k \rho_\alpha C^V_{\alpha , a}(\sigma )-\lambda_a\right).
\ee

In order to reproduce the precise formula given in section 1 we have to discuss the RSVW jacobian $J_{n}$.

\subsection{RSVW Jacobian}

At first sight, the RSVW equations are $2(n+k)$ equations
\be
\sum_{a=1}^n C^V_{\alpha ,a}(\sigma )\tilde\lambda_{a,\dot{\underline{\alpha}}}=0 ,\quad \sum_{\alpha=1}^k \rho_{\alpha,\underline{\alpha}} C^V_{\alpha , a}(\sigma ) = \lambda_{a,\underline{\alpha}}
\ee
for $2(n+k)$ variables
\be
\{ \rho_{1,1},\rho_{1,2},\ldots ,\rho_{k,1},\rho_{k,2}, \sigma_1^{(1)},\sigma_2^{(1)},\ldots ,\sigma_1^{(n)},\sigma_2^{(n)} \}
\ee
However, as noted by RSV \cite{Roiban:2004yf}, there are two problems with that statement which fortunately cancel each other. The first problem is that the equations when evaluated on external data, $\lambda_a$ and $\tilde\lambda_a$, satisfying momentum conservation become linearly dependent. In fact the rank is reduced from $2(n+k)$ down to $2(n+k)-4$. The second problem is that the set of equations is invariant under a $GL(2)$ action on the variables $\sigma_\alpha^{(a)}$ and $\rho_{\alpha,\underline\alpha}$. This seems to be a problem because using the $GL(2)$ action one can ``gauge fix" any four variables.

The nicest way to phrase both problems and later solve them is to construct the naive jacobian matrix pretending that this is a regular system of equations. In other words, construct a vector
\be
{\cal V} = \{ \rho_{1,1},\rho_{1,2},\ldots ,\rho_{k,1},\rho_{k,2}, \sigma_1^{(1)},\sigma_2^{(1)},\ldots ,\sigma_1^{(n)},\sigma_2^{(n)} \}
\ee
of variables and a vector
\be
{\cal E} = \{ E_{1,1},E_{1,2},\ldots ,E_{k,1},E_{k,2},F_{1,1},F_{1,2},\ldots ,F_{n,1},F_{n,2} \}
\ee
with
\be
E_{\alpha,\dot{\underline{\alpha}}} = \sum_{a=1}^n C^V_{\alpha ,a}(\sigma )\tilde\lambda_{a,\dot{\underline{\alpha}}},\quad F_{a,\underline{\alpha}} =\sum_{\alpha=1}^k \rho_{\alpha,\underline{\alpha}} C^V_{\alpha , a}(\sigma ).
\ee
Here $E_{\alpha,\dot{\underline{\alpha}}}=0$ and $F_{a,\underline{\alpha}}=\lambda_{a,\underline{\alpha}}$ are nothing but the RSVW equations.

The $2(n+k)\times 2(n+k)$ matrix is given by
\be
K_{I,J} = \frac{\partial {\cal E}_I}{\partial {\cal V}_J}.
\ee
Clearly, ${\rm det}(K) = 0$ and this reflects the problems mentioned above. This problem is quite reminiscent of the problem found by Hodges \cite{Hodges:2012ym} in his MHV computation. There too the construction starts with a singular matrix.

The key observation is that there is a canonical object that can be constructed from non-singular $2(n+k-2)\times 2(n+k-2)$ minors $K$ which is independent of the choice of minor made. In the introduction we mentioned a convenient one.

Eliminating any four rows corresponding to elements in $\cal V$ of the form $\{ \sigma_1^{(a)},\sigma_2^{(a)},\sigma_1^{(b)},\sigma_2^{(b)}\}$ and four columns corresponding to elements in ${\cal E}$ of the form $\{ F_{c,1},F_{c,2},F_{d,1},F_{d,2}\}$, we denote the non-singular minor by $K^{ab}_{cd}$. It turns out that the combination
\be
J_n = \frac{1}{(a~b)^2[c~d]^2} |K^{ab}_{cd}|
\ee
is independent of the choice of $\{ a,b,c,d\}$ and hence it is $S_n$ symmetric.

Of course, other choices are also possible but the complexity of the invariant form can change quite dramatically. We leave as an exercise for the reader to work out other choices.

Having concluded the derivation of the new formula the natural next step is to provide evidence for the orthogonality conjecture.

\section{Towards a \texorpdfstring{$G(k,n)$}{G(k,n)} Grassmannian Formulation}

While the formulation as an integral over $G(2,n)$ is very compact and in the case of Yang-Mills makes the KK and BCJ identities manifest, in practice the equations over the $\sigma$ variables are highly non-linear and hard to solve. Luckily, in Yang-Mills there is a way of converting the integral over $G(2,n)$ to one over $G(k,n)$ and the latter gives much simpler equations that can be solved using numerical methods to high precision. The technique is to integrate-in $GL(k)$ redundant variables describing a $k$-plane in ${\mathbb C}^n$ denoted as $C_{\alpha a}$ and impose constraints that force them to be $GL(k)$ equivalent to the Veronese map $C^V_{\alpha a}(\sigma )$. The last step is to integrate out the $\sigma$'s.

All checks of the orthogonality conjecture made in section 3 are actually carried out in the $GL(k)$ invariant formulation of the residues. It is therefore natural to try and write our formula in $GL(k)$ invariant form. We have succeeded in finding a simple form for $k=3$ and this is the main subject of this section.

\subsection{\texorpdfstring{$G(k,n)$}{G(k,n)} Grassmannian Formulation for SYM amplitudes}

Let us start with a short review of the connection between the RSVW formulation and the $G(k,n)$ formulation \cite{ArkaniHamed:2009dg, Nandan:2009cc}.

To establish the duality between the RSVW formula and the $G(k,n)$ Grassmannian formulation, an identity ``1'' is introduced in the connected formula:
\begin{equation}\label{eq:idTwistorString}
 1=\frac{1}{\text{Vol}(GL(k))}\int d^{k\times n}C_{\alpha a}\int d^{k\times k}L_{\alpha}^{\beta}(\text{det}(L))^n\prod_{\alpha,a}\delta\left(C_{\alpha a}-L_{\alpha}^{\beta}C_{\beta a}^V[\sigma]\right).
\end{equation}
The $\delta$-functions in this formulation ensure the localization of the variables $C$ to the Veronese form $C^V(\sigma)$ up to a $GL(k)$ transformation. This is the process of integrating-in $C$.

Integrating out the $G(2,n)$ variables and treating the remaining $\delta$-functions as poles yields
\begin{equation}\label{eq:TwistorString2}
\begin{split}
 \mathcal{T}_{n,k}=\frac{1}{\text{Vol}(GL(k))}&\int d^{k\times n}C_{\alpha a}d^{2k}\rho_\alpha\ \frac{H(C)}{S_1(C)\dots S_M(C)}\\ &\times \prod_{\alpha=1}^k\delta^{2}\big(C_{\alpha a}\tilde{\lambda}_a\big)\prod_{a=1}^n\delta^2\big(C_{\alpha a}\rho_\alpha-\lambda_a\big)\prod_{\alpha=1}^k\delta^{0|4}\big(C_{\alpha a}\tilde{\eta}_a\big)
\end{split}
\end{equation}
(Here we follow the convention in the literature \cite{ArkaniHamed:2009dg, Nandan:2009cc} of denoting the $GL(k)$ invariant formulation $\mathcal{T}_{n,k}$).

The integration is defined to compute the sum over all the residues associated to the isolated zeroes of the map $f:{\mathbb C}^M\to{\mathbb C}^M$ defined by the ``Veronese polynomials" $f=\{S_1,S_2,\ldots ,S_M\}$. Here $M=(k-2)(n-k-2)$ is determined by the number of integrations to be performed in the ACCK Grassmannian formulation. The general form of the Veronese polynomials for $k=3$ is known and reviewed below.

\subsubsection{NMHV: \texorpdfstring{$G(3,n)$}{G(3,n)} Grassmannian Formulation for Gauge Theory}

For general NMHV amplitudes each Veronese polynomials only depends on six particles at a time. This is why it is useful to explicitly show the labels and give the general definition
\begin{equation}\label{eq:S(C)}
 S(C)=S_{123456}=(123)(345)(561)(246)-(234)(456)(612)(135).
\end{equation}
Geometrically, the solutions of setting to zero all Veronese polynomials $S(C)$ constructed from all possible subsets of six labels of $\{ 1,2,\ldots , n\}$ are isolated (modulo $GL(k)$) and give rise to configurations of $n$ points,
$$\vec{C}_a = \left(
    \begin{array}{c}
      C_{a1} \\
      C_{a2} \\
      C_{a3} \\
    \end{array}
  \right) \in \mathbb{CP}^2,
$$
lying on a conic.

Of course, the set of all possible subsets of six elements has more elements than $M = (n-5)$ except for $n=6$ when both are $1$. How can one select only $n-5$ $S(C)$'s and still ensure that the only roots give all points on a conic? The answer is that it is not possible. For any set of $n-5$ $S(C)$ chosen there will be unphysical solutions, i.e., common roots where the $n$ points do not lie on a conic but accidentally set to zero the chosen set of Veronese polynomials \cite{Dolan:2009wf, ArkaniHamed:2009dg, Nandan:2009cc}. The resolution to this is simple. For any given choice there exists a function $H$ that vanishes on all the unphysical solutions. Since the ${\cal T}$ formula computes residues at the zeroes of $f$, having $H$ in the numerator ``selects" the physical solutions.

For the following set of Veronese polynomials
$$S_i=S_{1,2,3,i+3,i+4,i+5} \quad {\rm with} \quad i=1,\ldots ,n-5$$
one has
\begin{equation}\label{eq:H(C)}
H(C) = H_n(C)= \frac{\prod_{\mu=6}^{n-1}(1~2~\mu)(2~3~\mu\!-\!1)\prod_{\nu=5}^{n-1}(1~3~\nu)}{(1~2~3)(3~4~5)(n\!-\!1~n~1)}.
\end{equation}

\subsubsection{NMHV: \texorpdfstring{$G(3,n)$}{G(3,n)} Grassmannian Formulation for Gravity}

On the physical support of the Veronese conditions $S(C)=0$, the $C$ matrix can be brought to the $C^{V}$ form up to a $GL(k)$ transformation. It is therefore convenient to relate minors of $C$ to those of the $G(2,n)$ description directly without passing through the matrix representatives.

This can be easily done by realizing that up a $GL(1)$ rescaling
\begin{equation}
 (i~j~k)=-(i~j)(j~k)(k~i)
\end{equation}
and therefore the ratio
\begin{equation}
\label{trans}
 \left(\frac{(ij)}{(ab)}\right)^2=\frac{(a~i~j)(b~i~j)}{(i~a~b)(j~a~b)}
\end{equation}
becomes the relation we are looking for.

Let us start translating our $G(2,n)$ formula for gravity into $G(3,n)$ language. Consider the determinant in the numerator (\ref{andrew}), $H_n$, and the corresponding matrix $\Phi$ by choosing the reference $\sigma$'s to be equal $\ell = r$
\be
\Phi_{ab} = \left\{
         \begin{array}{cc}
           \frac{s_{ab}}{(a~b)^2} & {\rm for} \;\; a\neq b \\
           -\sum_{c=1, c\neq a}^n \frac{s_{ac}}{(a~c)^2}\frac{(c~r)^2}{(a~r)^2}  & {\rm for} \;\; a=b . \\
         \end{array}
       \right.
\ee
This is almost in the right form to be written in terms of $3\times 3$ minors. The last step is to define a new matrix $\widetilde\Phi$ by multiplying the $a^{\rm th}$ row of $\Phi$ by $(a+1~a+2)^2$. This gives a matrix that can be written in $G(3,n)$ form directly using (\ref{trans})
\be
\widetilde\Phi_{ab} = \left\{
         \begin{array}{cc}
           s_{ab}\frac{(a+1~a+2)^2}{(a~b)^2} & {\rm for} \;\; a\neq b \\
           -\sum_{c=1, c\neq a}^n s_{ab}\frac{(a+1~a+2)^2}{(a~c)^2}\frac{(c~r)^2}{(a~r)^2}  & {\rm for} \;\; a=b \\
         \end{array}
       \right.
\ee
Taking into account the changes introduced in the definition of the new matrix one finds
\begin{equation}
 H=c_{abc}c^{def}\big|\Phi^{(abc)}_{(def)}\big|=c_{abc}c^{def}\frac{(a\!+\!1~a\!+\!2)^2(b\!+\!1~b\!+\!2)^2(c\!+\!1~c\!+\!2)^2}{\left((1~2)(2~3)\dots(n~1)\right)^2}
 \big|\widetilde{\Phi}^{(abc)}_{(def)}\big|.
\end{equation}

An especially convenient choice, for reasons that will become apparent below, is given by $\{a~b~c\}=\{d~e~f\}=\{a~a\!+\!1~a\!+\!2\}$, yielding
\begin{equation}
 H=\frac{1}{\left((1~2)(2~3)\dots(n~1)\right)^2}\left(\frac{(a\!+\!2~a\!+\!3)}{(a~a\!+\!1)}\frac{(a\!+\!3~a\!+\!4)}{(a\!+\!2~a)}\right)^2\big|\widetilde{\Phi}^{(a~a\!+\!1~a\!+\!2)}_{(a~a\!+\!1~a\!+\!2)}\big|.
\end{equation}
Since $H$ is permutation invariant and the Parke-Taylor-like prefactor is cyclic invariant, the determinant including the $a$-dependent prefactor $\left(\frac{(a\!+\!2~a\!+\!3)}{(a~a\!+\!1)}\frac{(a\!+\!3~a\!+\!4)}{(a\!+\!2~a)}\right)^2\big|\widetilde{\Phi}^{(a~a\!+\!1~a\!+\!2)}_{(a~a\!+\!1~a\!+\!2)}\big|$ has to be cyclic invariant with respect to the ordering induced by the prefactor.

Using (\ref{trans}), the prefactor (and similarly $\widetilde{\Phi}$) can be rewritten in a manifestly $GL(3,n)$ invariant way in terms of $3\times 3$ minors:
\begin{equation}
\begin{split}
 \mathfrak{f}(a)&=\left(\frac{(a\!+\!2~a\!+\!3)}{(a~a\!+\!1)}\frac{(a\!+\!3~a\!+\!4)}{(a\!+\!2~a)}\right)^2\\
 &=\frac{(a~a\!+\!2~a\!+\!3)(a\!+\!1~a\!+\!2~a\!+\!3)}{(a\!+\!2~a~a\!+\!1)(a\!+\!3~a~a\!+\!1)}\frac{(a\!+\!2~a\!+\!3~a\!+\!4)(a~a\!+\!3~a\!+\!4)}{(a\!+\!3~a\!+\!2~a)(a\!+\!4~a\!+\!2~a)}.
\end{split}
\end{equation}

The final observation to getting a simple form for all NMHV amplitudes relating them to the square of their gauge theory counterpart is that the gauge theory answer can be written as
\be
A_n(1,2,\ldots ,n) = \sum_{i\in {\rm sol}(\rm RSV)}\left.\frac{1}{\left((1~2)(2~3)\dots(n~1)\right)}\times \frac{1}{J_n}\right|_{i}\delta^{0|4}\prod_{\alpha=1}^3\left( \sum_{a=1}^nC^V_{\alpha a}\tilde\eta_a \right).
\ee
This means that gravity amplitudes in the NMHV sector can be written as
\be
M_n = \sum_{I\in {\rm sol}({\rm RSV})} \mathfrak{f}(a)\big|\widetilde{\Phi}^{(a~a\!+\!1~a\!+\!2)}_{(a~a\!+\!1~a\!+\!2)}\big| \left(A^{(I)}_n(1,2,\ldots ,n)\right)^2
\ee
where the square of $\delta^{0|4}(C^V\tilde\eta)$ simply means to replace it by $\delta^{0|8}(C^V\tilde\eta)$.

\section{Evidence for the Orthogonality Conjecture}

In this section we provide numerical evidence for the orthogonality conjecture described in section 3. By directly computing the gauge theory residues of the RSVW formula for all amplitudes with eight or less particles and in all R-charge sectors we find that the conjecture holds.

As mentioned in the previous section, all computations of the RSVW residues are actually carried out in the $GL(k)$ invariant or ${\cal T}_{n,k}$ formulation.

Let us denote, as in section 3, the number of solutions to the RSV equations for $n$ particles in the $k$ R-charge sector by ${\cal N}_{n,k}$. No theoretical understanding of the structure of these number is available in the literature but a very natural guess can be made from the ones computed empirically. These are
\be
    \begin{array}{cccccc}
    n\setminus k  &  2 & 3 & 4 & 5 & 6 \\
      4 & 1 & \; & \; & \; & \; \\
      5 & 1 & 1 & \; & \; & \; \\
      6 & 1 & 4 & 1 & \; & \; \\
      7 & 1 & 11 & 11 & 1 & \; \\
      8 & 1 & 26 & 66 & 26 & 1 \\
    \end{array}
\ee
It is then natural to propose that in general ${\cal N}_{n,k}$ is the Eulerian number $\left\langle \!\! \begin{array}{c}
n-3 \\ k-2 \end{array} \!\! \right\rangle$. Eulerian numbers are obtained using the recursion relation
\be
\left\langle \!\! \begin{array}{c} p \\ q \end{array} \!\! \right\rangle = (p-q) \left\langle \!\! \begin{array}{c} p-1 \\ q-1 \end{array} \!\! \right\rangle + (q+1)\left\langle \!\! \begin{array}{c} p-1 \\ q \end{array} \!\! \right\rangle.
\ee

The orthogonality conjecture can be divided into two parts. The first is that all ${\cal N}_{n,k}$ residues associated to a given $n$ and $k$ are orthogonal among themselves using the KLT bilinear form. The second part is any individual residue in ${\cal N}_{n,k_1}$ is orthogonal to any individual residue in ${\cal N}_{n,k_2}$.

For each $n$ and $k$ we have computed numerically a ${\cal N}_{n,k}\times {\cal N}_{n,k}$ matrix corresponding to the KLT bilinear form in the residue basis. In other words, the $i,j$ entry is computed by
\be
\sum_{\alpha, \beta} {\cal I}^{\rm RSVW}_{i}(\alpha)\; S_{\alpha,\beta}^{\rm KLT}\;\; {\cal I}^{\rm RSVW}_j(\beta ).
\ee
We have found for all $n\leq 8$ and all $k\leq 4$ a diagonal matrix with very high numerical precision (see table below).

We have also computed the values of the KLT form for terms in different $k$ sectors and have found all zeroes with high precision.

For all numbers of particles, 20 calculations have been performed, with randomly generated external data satisfying only momentum conservation and using  WorkingPrecision at least 100 to find the roots of the polynomials\footnote{Here, WorkingPrecision 100 implies that the roots of the polynomials $P(\tau)$ are evaluated with a minimal accuracy of 100 digits.} As a reference for the orthogonality, the maximal cross-term (where the gauge theory amplitudes are evaluated on different solutions) is compared to the minimal diagonal term (where the gauge theory amplitudes are evaluated on the same solution). In the table below, the accuracy of the orthogonality relations is shown, defined as
\begin{equation*}
 \text{accuracy of the orthogonality}=\frac{|\text{maximal cross-term}|}{|\text{minimal diagonal term}|}.
\end{equation*}

Note that for the orthogonality between the single residues and the MHV or $\overline{\text{MHV}}$ amplitude, the highest order term is compared to the lowest order $k=3$ diagonal term. In the case when the gravity MHV amplitude (or $\overline{\text{MHV}}$) is smaller than the minimal diagonal term, this $\mathcal{N}=4$ SUGRA amplitude was used as a reference for the orthogonality.

The table below gives the mean accuracy, obtained by averaging over the data obtained from the 20 calculations for the $n=6,\ 7$ and $n=8$ point amplitudes and $k=3$. Included in the table is as well the variance from the average accuracy for each computation.

\begin{center}
\begin{tabular}{|c||c|c|c|} \hline \\[-10pt]
 number of &\multicolumn{3}{c|}{accuracy of the orthogonality in $\%$}\\
 particles &\multicolumn{3}{c|}{between the residues of $k=3$ and}\\[5pt]
 $n$ & MHV amplitudes & $\overline{\text{MHV}}$ amplitudes & residues of $k=3$\\[5pt] \hline  \hline & & & \\[-8pt]
 $n=6$ & $10^{-85\pm5}$ & $10^{-88\pm5}$ & $10^{-87\pm4}$\\
 $n=7$ & $10^{-72\pm5}$ & $10^{-77\pm4}$ & $10^{-75\pm5}$ \\
 $n=8$ & $10^{-36\pm6}$ & $10^{-42\pm6}$ & $10^{-30\pm5}$ \\ \hline
\end{tabular}
\end{center}

For $n=8$ and $k=4$ we find that all $\mathcal{N}_{8,4}=66$ solutions are orthogonal to one another with an accuracy of $10^{-150}$. Furthermore, all 66 solutions are orthogonal to the MHV and $\overline{\text{MHV}}$ amplitudes as well as the 26 $k=3$ solutions with similarly high accuracies. This provides a highly non-trivial check on the both parts of the orthogonality conjecture: All $\mathcal{N}_{8,k}$ residues for $k\in\{2,3,4,5,6\}$ are orthogonal among themselves, and the calculation of the N$^2$MHV residues confirmes the orthogonality of the single residues in different sectors $k_1$ and $k_2$.

\subsection{KLT Bilinear Form}

One of the well-known properties of the Eulerian numbers $\left\langle \!\! \begin{array}{c}
 p\\ q \end{array} \!\! \right\rangle$ is that for fixed $p$ the sum over all $q$'s gives $p!$. In our case this means that the sum of ${\cal N}_{n,k}$ for fixed $n$ and over all $k$'s equals $(n-3)!$.

Before 2008, the basis of partial amplitudes in Yang-Mills theory for a fix $n$ was expected to have $(n-2)!$ elements after using the KK relations. In 2008, BCJ showed that the basis can be vastly reduced down to $(n-3)!$. This is quite interesting as it shows that when using a particular choice of KK-BCJ basis for partial amplitudes, the KLT bilinear form becomes a $(n-3)!\times (n-3)!$ matrix.

Assuming the validity of the orthogonality conjecture and of the relation ${\cal N}_{n,k}=\left\langle \!\! \begin{array}{c} n-3 \\ k-2 \end{array} \!\! \right\rangle$ one concludes that the RSVW residues form a complete orthogonal basis with respect to the KLT bilinear form.

This connection between such different ways of getting $(n-3)!$, one in gauge theory and the other in gravity, lead us to believe that a proof of the orthogonality conjecture will reveal a very interesting fact about the connection between the two theories. At the moment, the best lead comes from the fact that all ${\cal N}_{n,k}$ residues satisfy both the KK and BCJ relations independently \cite{Roiban:2004yf, newpaper} and we believe that this is the most promising direction towards a proof.

\section{Future Directions}

We have provided strong evidence for a new formulation of the tree-level S-matrix of ${\cal N}=8$ supergravity which can be thought of as the analog of the connected or RSVW formulation of ${\cal N}=4$ super Yang-Mills. In this final section we discuss some of the natural questions that arise and which, in our view, will be interesting to pursue.

\subsection{Off-Shell MHV Amplitudes: Correlation Functions}

In gauge theory the core of the RSVW formula
\be
\frac{1}{(1~2)(2~3)\ldots (n~1)}
\ee
has exactly the same structure as the Parke-Taylor form of MHV amplitudes. In \cite{Nair:1988bq}, Nair gave a beautiful interpretation of the Parke-Taylor formula as a correlation function on a $\mathbb{CP}^1$. In Witten's construction \cite{Witten:2003nn} of the twistor string, the RSVW formula can be thought of as the Veronese map into twistor space of the world-sheet correlation function.

The formula for gravity presented in this paper shares similar features. In particular, at its core one starts with a novel off-shell version of MHV amplitudes. We have shown that the KLT MHV formula and Hodges MHV formula are equivalent after the off-shell deformation. The way we achieved that is by showing that they satisfy the same recursion relations. It would be very interesting to show that all other MHV formulas known in the literature satisfy the same off-shell version of Hodges' recursion relation. This would then hint at a correlation function interpretation of the off-shell formula along the lines perhaps of Nair \cite{Nair:2005iv} or Mason-Skinner \cite{Mason:2008jy}.

\subsection{Manifestly Permutation Invariant Form: Fermionic and Bosonic Redundancies}

As usual in physical set ups, the price for making a symmetry manifest is the introduction of redundancies. The quest for a manifestly symmetric form of tree level gravity amplitudes proved to be a difficult problem already for one of the simplest amplitudes, the four graviton amplitude. All known formulas do not seem to be $S_4$ invariant, in fact, they are only so on the support of the momentum conserving delta function and usually require some manipulations to prove it. The question is then what redundancy has to be introduced for gravity amplitudes.

A beautiful way to write down the four particle amplitude in a manifestly $S_4$ invariant way was obtained in \cite{nima} by introducing a formula with a fermionic redundancy. The known formulas are obtained after gauge fixing the redundancy which has to be done by making some choices that single out some particles from the rest.

It is natural to ask whether the same can be applied to our gravity formula. Let us now see that this is indeed the case.

The first step is to relate the numerator $H_n$ to the following manifestly $S_n$ invariant fermionic integral
$$ Z_n = \frac{1}{{\rm vol}(G_F)}\int \prod_{a=1}^nd\chi_a d\eta_a \exp \left( \chi_a \Phi_{ab}\eta_b \right).$$
Here ${\rm vol}(G_F)$ is the volume of the fermionic redundancy and it is zero while $\{\chi_a,\eta_a\}$ are Grassmann variables.

In order to compute any explicit amplitude we have to gauge fix using the Faddeev-Popov procedure. The matrix $\Phi_{ab}$ has rank $n-3$ and therefore the dimension of its null space is $3$.

Let's denote by $\{ v^{(1)},v^{(2)},v^{(3)}\}$ and $\{ {\tilde v}^{(1)},{\tilde v}^{(2)},{\tilde v}^{(3)}\}$ two sets of basis of the null space of $\Phi$. The redundancy in $Z$ is given by
$$ \eta_i \to \eta_i + \sum_{a=1}^3 v^{(a)}_i \theta_a, $$
$$ \chi_i \to \chi_i + \sum_{a=1}^3 {\tilde v}^{(a)}_i \tilde\theta_a .$$
Applying the gauge fixing leads to
$$ Z_n = c_{i_1,i_2,i_3}{\tilde c}_{j_1,j_2,j_3}\int \!\!\! \prod_{I=1,I\nin \{ i_1,i_2,i_3\}}^n \prod_{J=1,J\nin \{j_1,j_2,j_3\}}^n \!\!\!\!\! d\chi_I d\eta^J \exp\left( \sum_{I=1,I\nin \{ i_1,i_2,i_3\}}^n\sum_{J=1, J\nin \{j_1,j_2,j_3\}}^n \!\!\!\!\! \chi_I \Phi^I_J\eta^J \right).$$
where $c$ and $\tilde c$ come from the Faddeev-Popov jacobians and coincide with the definition given in section 1. We then conclude that $Z_n = H_n$.

The other ingredient needed is to write the denominator $J_n$ in the gravity formula in a manifestly $S_n$ invariant form. This is actually achieved in a much more conventional way. The idea is the same but introducing instead a bosonic redundancy.

It is possible to show that $1/J_n$ is
\be
\label{boso}
\frac{1}{{\rm vol}(G_B)}\int d\phi_{\alpha,\underline{\dot\alpha}}d\phi_{\alpha,\underline{\dot\alpha}}^*
d\varphi_{a,\underline{\alpha}}d\varphi_{a,\underline{\alpha}}^* {\rm exp}\left( \phi^{*T} \left(\frac{\partial E}{\partial \sigma}\right)\varphi + \varphi^{*T} \left(\frac{\partial F}{\partial \rho}\right)\phi + \varphi^{*T} \left(\frac{\partial F}{\partial \sigma}\right)\varphi \right)
\ee
where $G_B$ is the bosonic redundancy coming from the fact that the matrix ${\cal K}$ defined in section 1 has co-rank $4$.
It would be very interesting to find a way of combining both redundancies. The main obstacle is that the number of fermionic and bosonic redundancies is different. One possible way to overcome this problem is by stripping out a bosonic rescaling redundancy.

\subsection{Twistor Space}

The RSVW formulation originated in Witten's twistor string construction \cite{Witten:2003nn} and it has a beautiful representation in super-twistor or dual super-twistor space
\be
{\tilde {\cal A}}^{(k)}(1,2,\ldots , n) = \frac{1}{{\rm vol}(GL(2))}\int \frac{d^{2n}\sigma}{(1~2)(2~3)\ldots (n~1)}\prod_{\alpha=1}^k\delta^{4|4}\left(\sum_{a=1}^n C^V_{\alpha ,a}(\sigma ){\cal W}_a \right).
\ee
It is natural to ask for a twistor space representation of the gravity amplitude. At this point it is clear that in the form of the ratio of two determinants it is very hard to determine the twistor space formulation. However, if we use the integral with bosonic redundancy introduced above then it is possible to get a somewhat interesting formula. The key is to notice that only one of the three terms in the exponent of (\ref{boso}) affects the half-Fourier transform into twistor space. The relevant modification is
\be
\int\prod_{a=1}^nd^2\lambda_a e^{i\sum_a\ab{\lambda_a~\mu_a}}\int \prod_{\alpha=1}^kd^2\rho_\alpha \prod_{b=1}^n\delta^2(\rho_\alpha C^V_{\alpha a}(\sigma ) -\lambda_a)\; {\rm exp}\left(\sum_{a=1}^n\phi^*_{a\underline{\alpha}}\sum_{\alpha=1}^k\rho_{\alpha\underline{\alpha}}\frac{\partial C^V_{\alpha a}}{\partial \sigma^{(a)}_{\underline{\beta}}}\phi_{a\underline{\beta}}\right)
\ee
As usual, the delta functions are used to carry out all $\lambda$ integrations. The new piece is the exponential which depends on $\rho$ and therefore modifies the delta functions produced when the integrations over $\rho$ are performed.

The other significant change compared to the gauge theory twistor space answer is that the numerator in the integrand, i.e., $H_n$, becomes a differential operator, ${\widehat H}_{n}$.

The final form of the twistor space version of the gravity formula is
\be
{\tilde {\cal M}}^{(k)}_n = & \frac{1}{{\rm vol}(GL(2))}\int d^{2n}\sigma {\widehat H}_{n} \frac{1}{{\rm vol}(G_B)}\int d\phi_{\alpha,\underline{\dot\alpha}}d\phi_{\alpha,\underline{\dot\alpha}}^*
d\varphi_{a,\underline{\alpha}}d\varphi_{a,\underline{\alpha}}^* {\rm exp}\left( \phi^T \left(\frac{\partial E}{\partial \sigma}\right)\varphi + \varphi^T \left(\frac{\partial F}{\partial \rho}\right)\phi\right)\times \nonumber \\ & \prod_{\alpha=1}^k \delta^{0|8}(C_{\alpha a}^V\tilde\eta_a)\delta^{2}(C_{\alpha a}^V\tilde\lambda_a)\delta^{2}(C_{\alpha a}^V\mu_a + \phi^*_{a\underline{\alpha}}\frac{\partial C^V_{\alpha a}}{\partial \sigma^{(a)}_{\underline{\beta}}}\phi_{a\underline{\beta}})
\ee
where the differential operator ${\widehat H}_{n}$ is obtained from $H_n$ by the replacement
\be
s_{ab} = \ab{a~b}[a~b] \rightarrow -[a~b]\ab{\frac{\partial}{\partial \mu_a}~\frac{\partial}{\partial \mu_b}}.
\ee
It is interesting that the breaking of conformal invariance has two different origins. The first is the differential operator ${\widehat H}_{n}$ which is the one responsible for the derivative-of-a-delta function support of gravity amplitudes first noticed in \cite{Witten:2003nn}. The second is some sort of deformation of the localization of $\mu$. While in gauge theory one has that $C_{\alpha,a}^V\mu_a = 0$ in gravity the right hand side is deformed. It would be very interesting to find out the geometric meaning of these deformations.

\section*{Acknowledgments}

We thank R. Boels, Y. Gu, and S. Rajabi for useful discussions. F.C. thanks B. Feng and S. He for explaining their work on KLT and BCJ. F.C. also thanks N. Arkani-Hamed numerous discussions and encouragement. The authors also thank J. Bourjaily for sharing some gauge theory numerical data. This research was supported in part by the NSERC of Canada and MEDT of Ontario.

%\bibliographystyle{nhieeetr}
%\bibliography{refs.bib}

\end{document}